\documentclass{article}

\usepackage{arxiv}
\usepackage[utf8]{inputenc} 
\usepackage[T1]{fontenc}    
\usepackage{hyperref}       
\usepackage{url}            
\usepackage{booktabs}       
\usepackage{amsfonts}       
\usepackage{nicefrac}       
\usepackage{microtype}      
\usepackage{lipsum}
\usepackage{graphicx}
\usepackage{xcolor}
\usepackage{todonotes}
\usepackage{comment}
\usepackage[super]{nth}
\usepackage{amssymb}
\usepackage{amsmath}
\usepackage{bm}
\usepackage{commath}
\usepackage{subcaption}
\usepackage{siunitx}
\usepackage{cleveref}
\graphicspath{ {./images/} }
\renewcommand{\vec}[1]{\bm{#1}}
\newcommand{\tin}{\text{in}}

\title{Direct Numerical Simulation of a Turbulent Boundary Layer on a Flat Plate Using Synthetic Turbulence Generation}

\author{
 James Wright \\
  Ann and H. J. Smead Aerospace Engineering Sciences\\
  University of Colorado Boulder\\
  Boulder, CO 80309 \\
  \texttt{james.wrightiii@colorado.edu} \\
  \And
  Riccardo Balin \\
  Ann and H. J. Smead Aerospace Engineering Sciences\\
  University of Colorado Boulder\\
  Boulder, CO 80309 \\
  \texttt{riccardo.balin@colorado.edu} \\ 
  \And
  John W. Patterson \\
  Ann and H. J. Smead Aerospace Engineering Sciences\\
  University of Colorado Boulder\\
  Boulder, CO 80309 \\
  \texttt{jopa@colorado.edu} \\ 
  \AND
  John A. Evans \\
  Ann and H. J. Smead Aerospace Engineering Sciences\\
  University of Colorado Boulder\\
  Boulder, CO 80309 \\
  \texttt{john.a.evans@colorado.edu} \\
   \And
 Kenneth E. Jansen \\
  Ann and H. J. Smead Aerospace Engineering Sciences\\
  University of Colorado Boulder\\
  Boulder, CO 80309 \\
  \texttt{jansenke@colorado.edu} \\
}

\begin{document}
\maketitle

\begin{abstract}
The turbulent boundary layer over a flat plate is computed by direct numerical simulation (DNS) of the incompressible Navier-Stokes equations as a test bed for a synthetic turbulence generator (STG) inflow boundary condition. 
The inlet momentum thickness Reynolds number is approximately \num{1000}. 
The study provides validation of the ability of the STG to develop accurate turbulence in 5 to 7 boundary layer thicknesses downstream of the boundary condition. 
Also tested was the effect of changes in the stabilization scheme on the development of the boundary layer.
Moreover, the grid resolution required for both the development region and the downstream flow is investigated when using a stabilized finite element method.
\end{abstract}



\section{Introduction}
\label{sec:Intro}

Direct numerical simulation (DNS) resolves all of the turbulent scales of motion leaving none to be modeled on the interior of the domain.  
The only modeling approximation is introduced through its need for unsteady boundary conditions.
For some flow conditions, periodicity can be employed, though this limits the largest length scale that may influence the solution.  
For spatially developing flows, the influence of the boundary condition is felt more directly through the prescription of an unsteady inflow boundary condition. 
This paper studies the influence of a particular synthetic turbulence generator (STG) boundary condition proposed in \cite{Shur_STG}. 

This paper is organized as follows. Section~\ref{sec:NumApproach} describes in detail the flow problem chosen and the numerical approach taken to obtain a solution with DNS.
Section~\ref{sec:Results} presents and discusses the results obtained from the DNS. 
Finally, Section~\ref{sec:Conc} offers some concluding remarks.

\section{Numerical Setup}
\label{sec:NumApproach}

\subsection{Problem Definition}
\label{sec:NumApproachDef}
The flow computed in this study is the turbulent boundary layer over a zero pressure gradient flat plate. The momentum thickness Reynolds number at the inflow is $Re_\theta=970$, which is typical of DNS of more complex flows \cite{BalinAviation2020}.  
While far from the highest $Re$ simulated \cite{Schlatter2009,Jimenez_ZPGDNS,Sillero_DNS_2013}, the goal here is different.  By choosing a modest size simulation, a careful study of the inflow boundary condition and other aspects of the simulation, such as numerical method (and associated dissipation) and the grid resolution requirements, can be compared.  
\subsection{Solution Approach}
\label{sec:NumApproachSol}
At the inflow, the STG method of \cite{Shur_STG} was selected in order to provide unsteady and spatially varying velocity fluctuations into the domain. This method has been shown to produce realistic turbulence a short distance downstream of the inlet for both wall-modeled LES and DNS \cite{Shur_STG,Spalart_BJWMLES}. 
According to this approach, velocity fluctuations are first computed from a superposition of spatiotemporal Fourier modes with random amplitudes and phases. These are then scaled by prescribed profiles of the time-averaged Reynolds stresses in order to obtain the desired second order moments and thus introduce synthetic scales with both anisotropy and inhomogeneity. The fluctuations are finally added to a known mean velocity profile to obtain a time and spatially varying boundary condition. 

While the original concept of STG dates back to \cite{Kraichnan70}, more detail on the history, development, and analysis of the STG method can be found in \cite{Shur_STG} and in \cite{Patterson2020}. Here we summarize the equations and description from \cite{Patterson2020}.

\begin{equation}
u'_i=a_{ij}2\sqrt{\frac{3}{2}}\sum_{n=1}^N\sqrt{q^n}\sigma^n_j\cos(\kappa^nd^n_l\hat{x}_l+\psi^n) 
\label{eqn:uPrime}
\end{equation} 
\begin{center}
\begin{tabular}{l l l}
     $\hat{x}_i\equiv\{{(x_1-U_0 t')}\max(\kappa_{e}^{min}/\kappa^n,0.1),x_2,x_3\}$&
      $\sigma_j^nd^n_j=0$ &
     $a_{ik}a_{jk}=R_{ij}$ \\ 
     $q^n\equiv\int_{\kappa^{n-1}}^{\kappa^n}E(\kappa^n)d\kappa/\int_0^{\infty}E(\kappa^n)d\kappa$&
     $\sigma^n=\sigma^n(\theta^n,\phi^n)$& \\
     $E(\kappa^n)=(\kappa^n/\kappa_e)^4 f_\eta f_{cut} [1+2.4(\kappa^n/\kappa_e)^2]^{-17/6}$ &
     $d_i^n=d_i^n(\theta^n,\phi^n,\eta^n)$\\

     \end{tabular}
     \newline
\end{center}
     
Einstein summation notation has been adopted when the contraction is between spatial dimensions (subscripts) while a sum is written for Fourier components (superscripts) since they are often repeated more than twice.
The terms $\theta^n$, $\phi^n$, $\eta^n$, and $\psi^n$ are sets of random variables defined by their probability density functions and intervals: $f_{\theta}=\sin(\theta)/2$; $\theta \in [0,\pi]$, $f_{\phi,\eta,\psi}=\pi/2$; $\{\phi,\eta,\psi\} \in [0,2\pi)$. 
The two random sets of spherical angles $\theta^n$ and $\phi^n$ cause the set of unit vectors, $\sigma_j^n$, to be uniformly distributed on a unit sphere.
Imposing a divergence-free velocity ($\sigma_j^nd^n_j=0$ as verified in \cite{Smirnov01}) together with the requirement that $d_j^n$ 
be uniformly distributed on a unit sphere leaves $d_j^n$ a function of $\sigma_j^n$'s dependent variables ($\theta,\phi$) and the angle $\eta$ in the plane normal to $\sigma_j^n$. The random angles, radial lengths, and defined intensities are wave modes that have been mapped to wave space from a pseudo isotropic turbulence via the spatial Fourier transform. Note, the definition of $\hat{x}_i$  provides a means to slide through this pseudo turbulence domain by a spatial coordinate that progresses at the inflow bulk speed, $U_0$, naturally accounting for bulk convection. The second argument of the $\max$ is due to \cite{ShurPC}. Definitions for $\kappa_e^{min}, f_\eta, f_{cut}, k_e(x_i)$ match those given in \cite{Shur_STG} where they were chosen to adjust the spectrum (eddy size and anisotropy) to accurately account for proximity to solid boundaries. Further, their approach of defining $a_{ij}$ through a Cholesky decomposition of the Reynolds stress, $R_{ij}$, is adopted.

The mean velocity and Reynolds stress profiles for the flow analyzed here were obtained from Case C of the DNS study of \cite{Coleman_DNS_2018} at $Re_\theta=1,300$ and rescaled to the lower $Re_\theta=970$ using the relations in \cite{LundWuSquires_inflowRecycling}. 
Though it seem intuitive that normal stress profiles obtained directly from the DNS would be the best choice in order to inject synthetic scales with the correct anisotropy, in our experience and that of colleagues 
isotropic normal stress profiles perform better for this STG method. These are computed from the Reynolds shear stress as outlined in Eq.~\ref{eq:SATKE} in a similar manner to what would be performed with the solution of a RANS simulation with the Spalart-Allmaras one-equation model \cite{spalart1994one}. Note that $k^\text{mod}$ is the modeled turbulent kinetic energy and $C_\mu=0.09$.
\begin{equation}
    k^\text{mod} = \frac{-\overline{u'v'}}{C_\mu}
    \qquad
    \qquad
    \overline{u'u'}=\frac{2}{3}k^\text{mod}=\overline{v'v'}=\overline{w'w'}
    \label{eq:SATKE}
\end{equation}


Other boundary conditions of the DNS were as follows. 
The plate was treated as a no-slip wall with zero velocity. 
The top surface was tilted slightly downward so that it could be treated like the outflow where weak enforcement of zero pressure was applied along with zero traction. At the inflow and outflow, the top boundary was located $21\delta_{\text{in}}$ and $20\delta_{\text{in}}$ above the flat plate, respectively, where \(\delta_\text{in}\) is the boundary layer thickness at the domain inlet.
The outflow was located $27\delta_{\text{in}}$ downstream of the inflow where weak enforcement of zero pressure and traction were also applied. Effects from this boundary condition on the interior domain were contained within a streamwise distance of one local boundary layer thickness and thus did not affect the upstream solution. 
The spanwise period was set to $3\delta_{\text{in}}$ with periodicity enforced.  

Two structured hexahedral grids were constructed for the simulation. 
First was a baseline grid with spacing $\Delta x^+=15$, $\Delta z^+=6$, $\Delta y^+_1=0.1$, and $\Delta y^+_\text{max}=10$, where the inflow friction velocity was used for the non-dimensionalization. 
This resulted in a total of 31 million points.
Second was a coarse grid, where the \(\Delta x^+\) and \(\Delta z^+\) spacings were doubled to 30 and 12, respectively, while the \(\Delta y^+\) spacing was maintained from the baseline mesh.

The simulations were integrated for at least five domain flow-through times, which corresponds to at least 80 eddy turnover times in the domain, defined as the local edge velocity divided by the boundary layer height. 
This was deemed statistically stationary by looking at various sub-windows of the total accumulated time average.

\subsection{Flow Solver}
\label{sec:NumApproachCode}
The Parallel-Hierarchic-Adaptive-Stabilized-Transient-Analysis (PHASTA) flow-solver was used for all simulations presented in this work. PHASTA uses a stabilized, semi-discrete finite element method to solve either the compressible or incompressible Navier-Stokes equations plus any additional set of 
scalar equations for turbulence modeling. 
While high-order hierarchical bases of polynomials are available, tri-linear hexahedral elements were selected for these simulations resulting in second order accuracy.
Stabilization is performed with the streamline upwind/Petrov-Galerkin (SUPG)
method \cite{Whiting1999}, described in \cref{sec:AppSUPG}.
Time integration is performed with the fully implicit, second-order accurate generalized-$\alpha$ method \cite{Jansen_GenAlpha_2000}. 
These high-fidelity, massively parallel computations could be performed in a reasonable time frame due to the strong scalability of PHASTA, 
which has been demonstrated in \cite{Sahni_PHASTAScaling_09}.
The accuracy of DNS with PHASTA has been shown for a channel flow in \cite{Trofimova_DNS_2009}, in which tri-linear hexahedral elements were also used. 
Note finally that the incompressible branch of the solver was selected for this work due to the low Mach number of the flow.

\section{Results}
\label{sec:Results}

\subsection{Baseline and Coarse DNS Results}\label{sec:ResBaseDNS}
\begin{figure}
\centering
  \includegraphics[width = 0.6\textwidth]{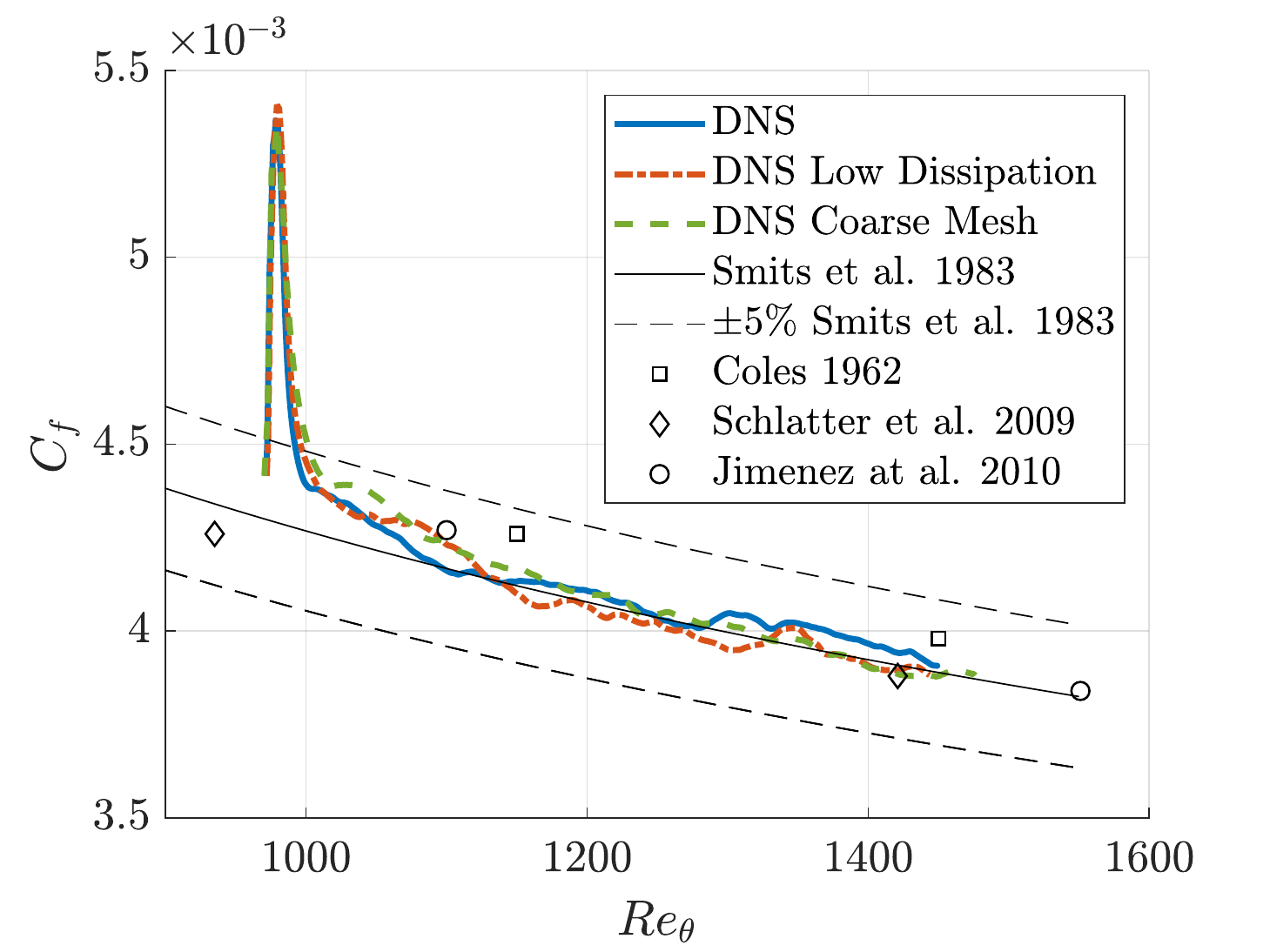}
  \caption{Comparisons of skin friction coefficient for a zero pressure gradient flat plate DNS using two different dissipation settings.}
\label{fig:DisComp_Cf}
\end{figure}

The skin friction coefficient computed from the baseline grid is shown by the solid curve 
in \cref{fig:DisComp_Cf} along with data from other DNS and experiments \cite{Coles1962,Schlatter2009,Smits1983,Jimenez_ZPGDNS}. 
After an expected initial spike immediately downstream of the STG inflow caused by the development of physically realistic near-wall turbulent structures, the $C_f$ quickly drops and good agreement with the validation data is obtained. 
The coarse mesh \(C_f\) result, also presented in \cref{fig:DisComp_Cf}, is shown to agree well with the baseline mesh and data from other DNS and experimental correlations.

\begin{figure}
     \centering
     \begin{subfigure}[b]{\textwidth}
         \centering
          \includegraphics[width = 0.33\textwidth]{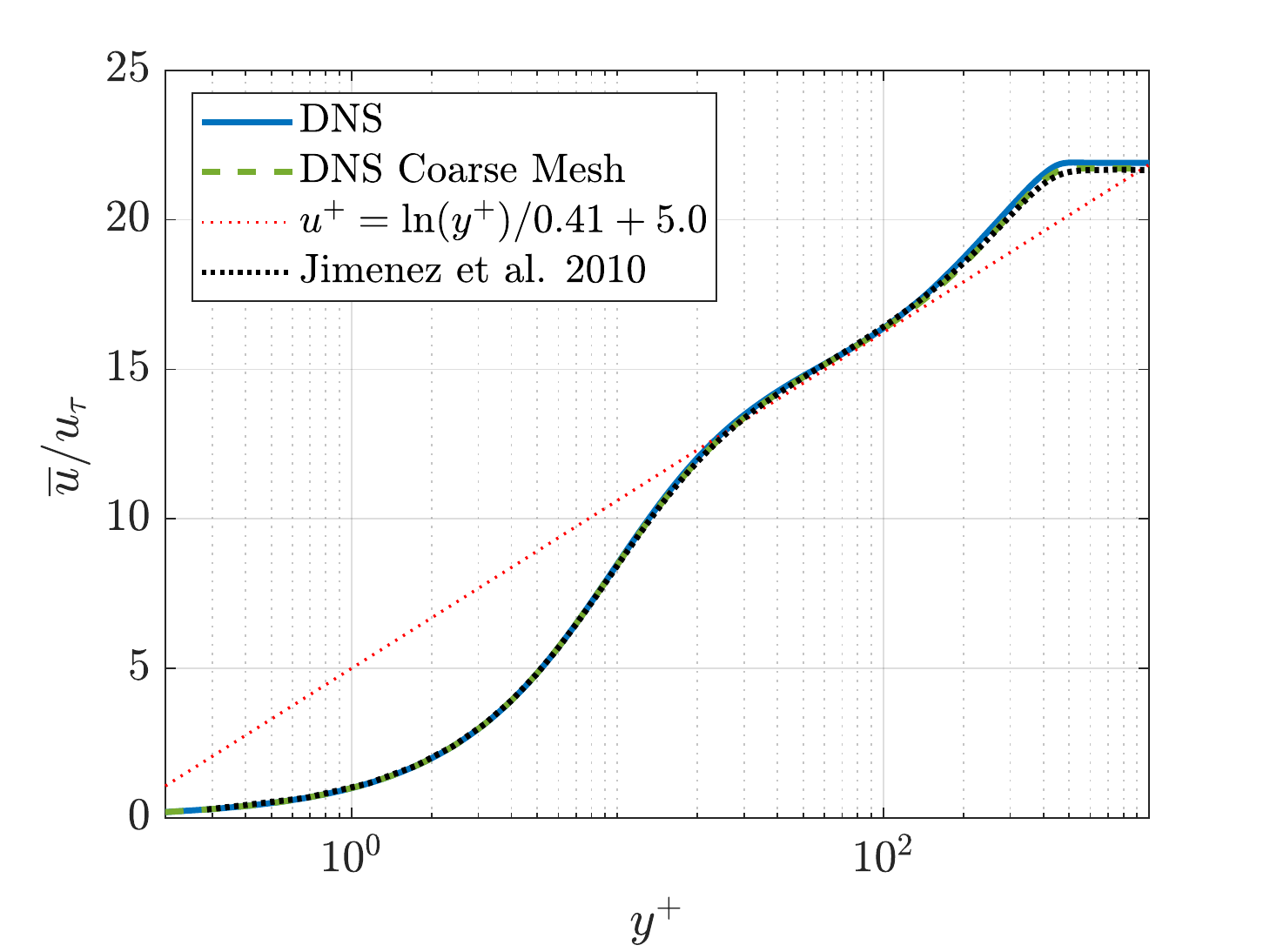}
          \includegraphics[width = 0.33\textwidth]{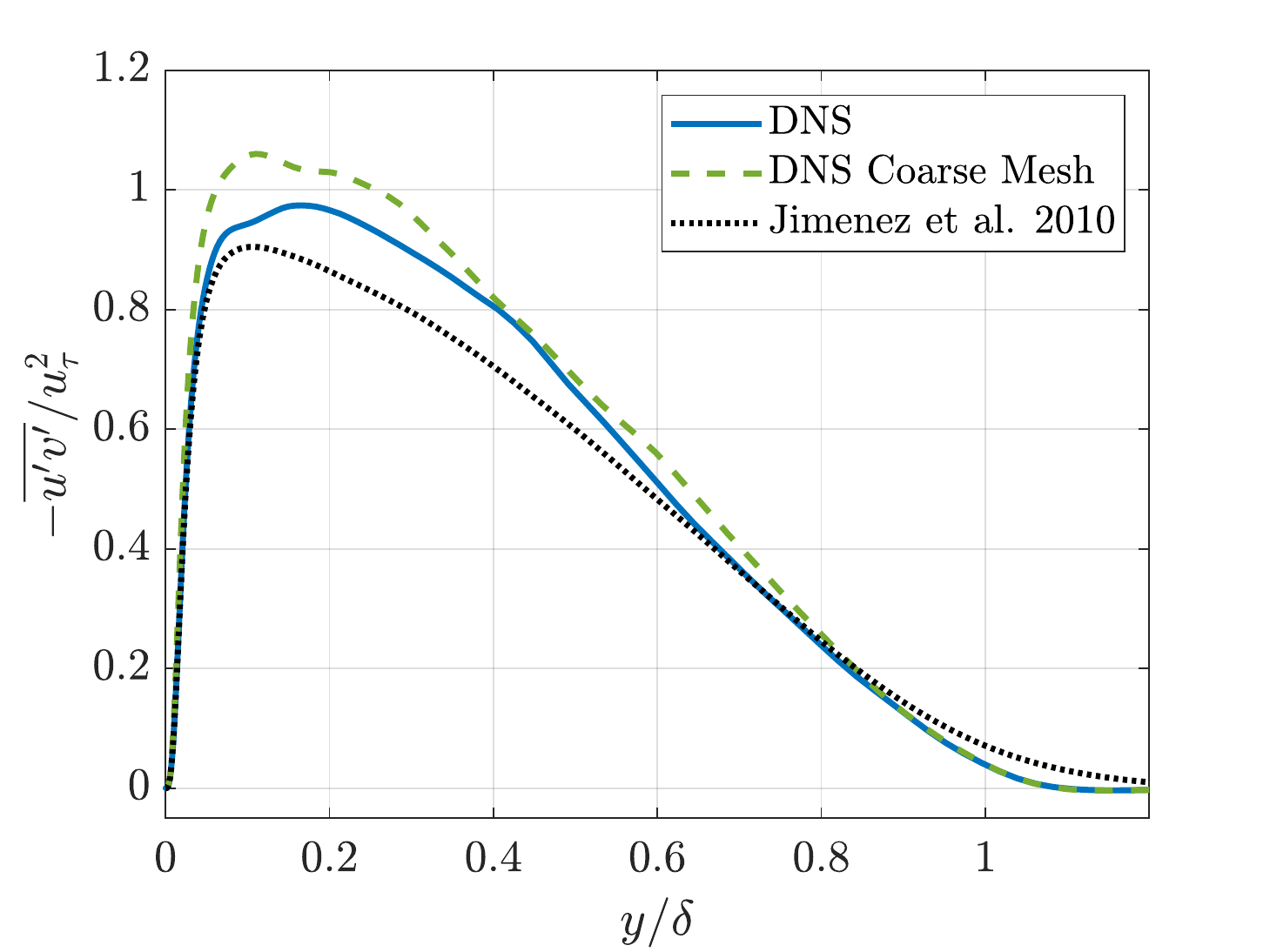}
          \includegraphics[width = 0.33\textwidth]{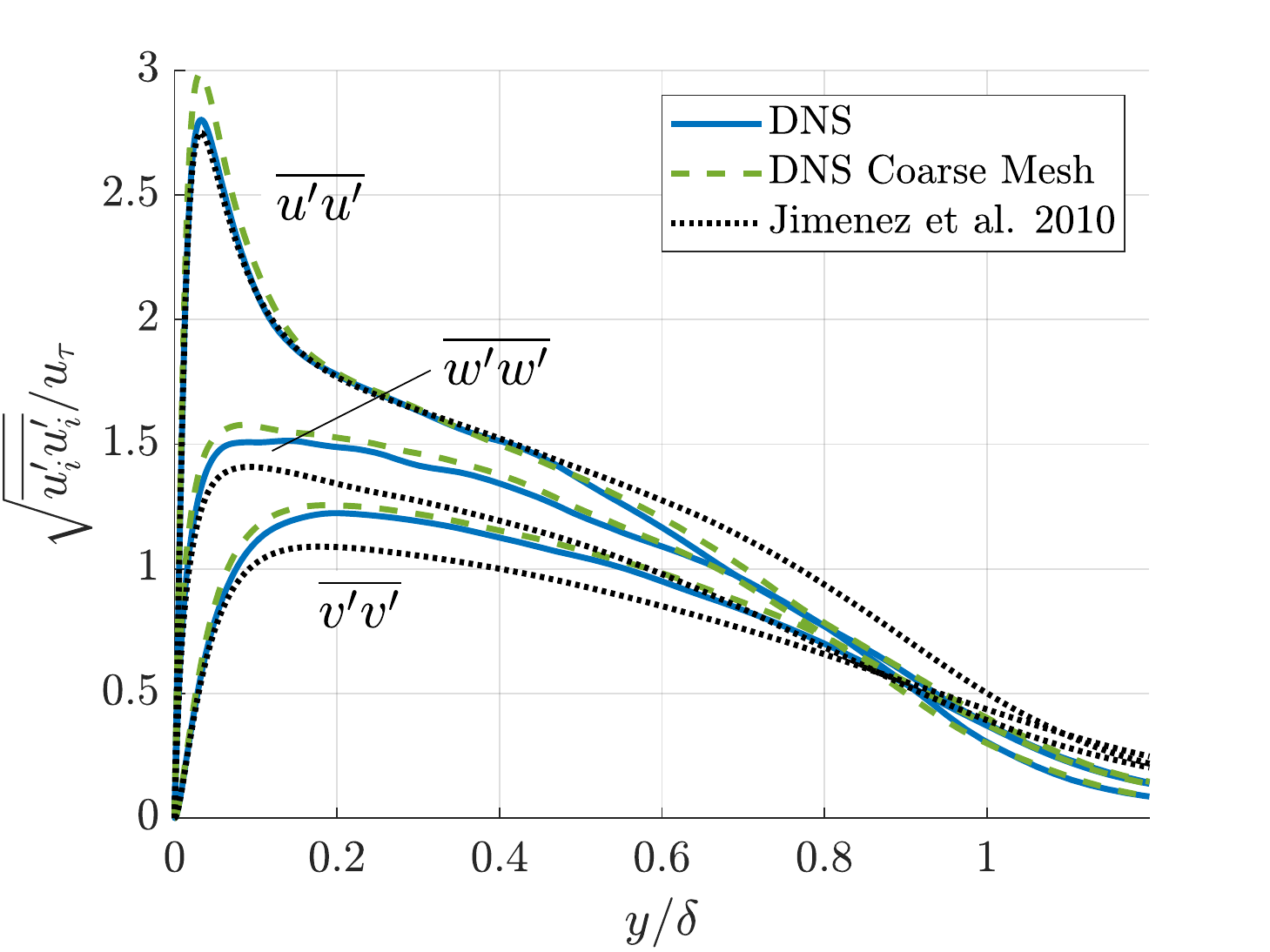}
         \caption{\(Re_\theta = \num{1100}\)}
         \label{fig:GridComp_Re1100}
     \end{subfigure}
     \begin{subfigure}[b]{\textwidth}
         \centering
          \includegraphics[width = 0.33\textwidth]{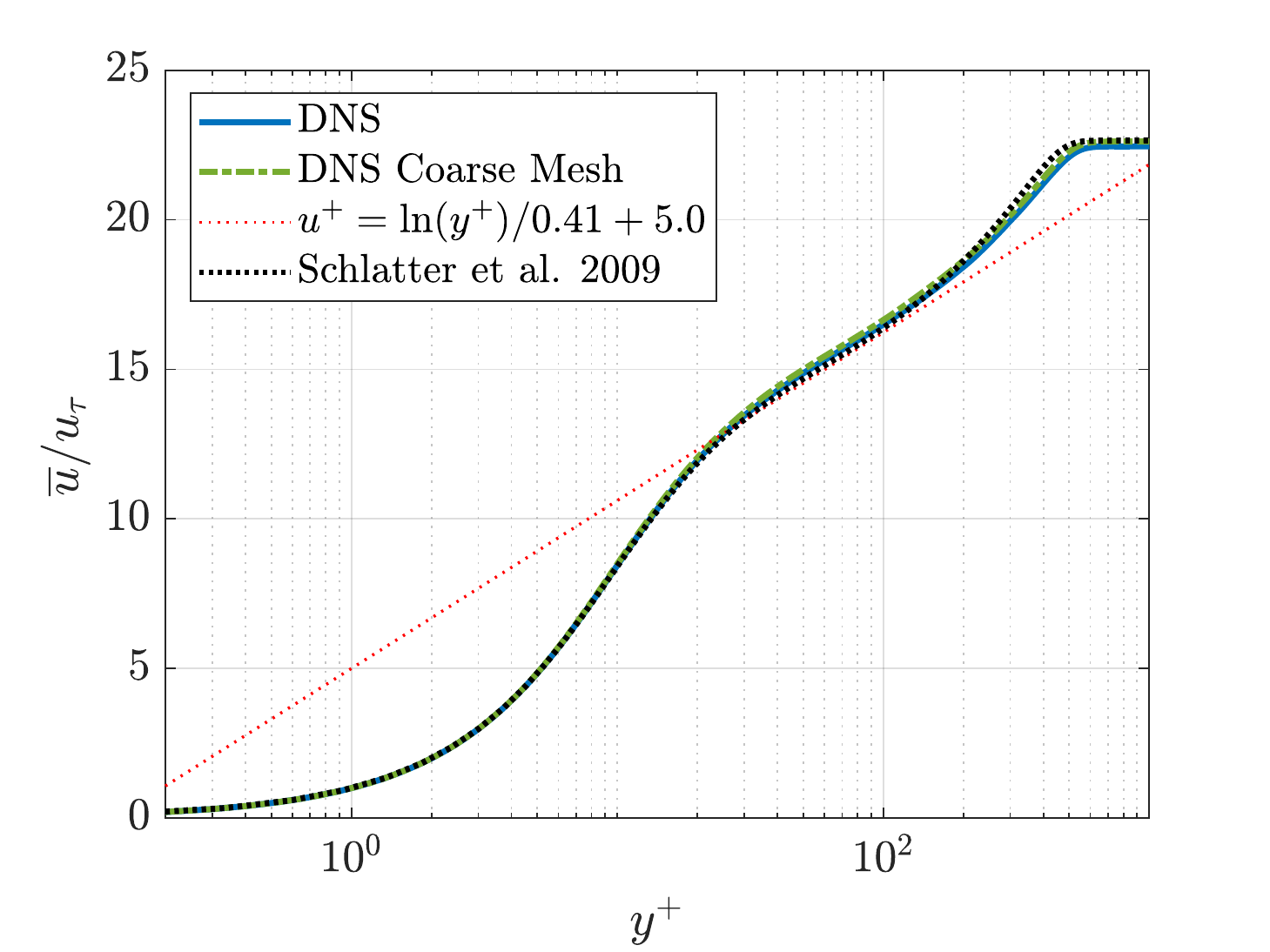}
          \includegraphics[width = 0.33\textwidth]{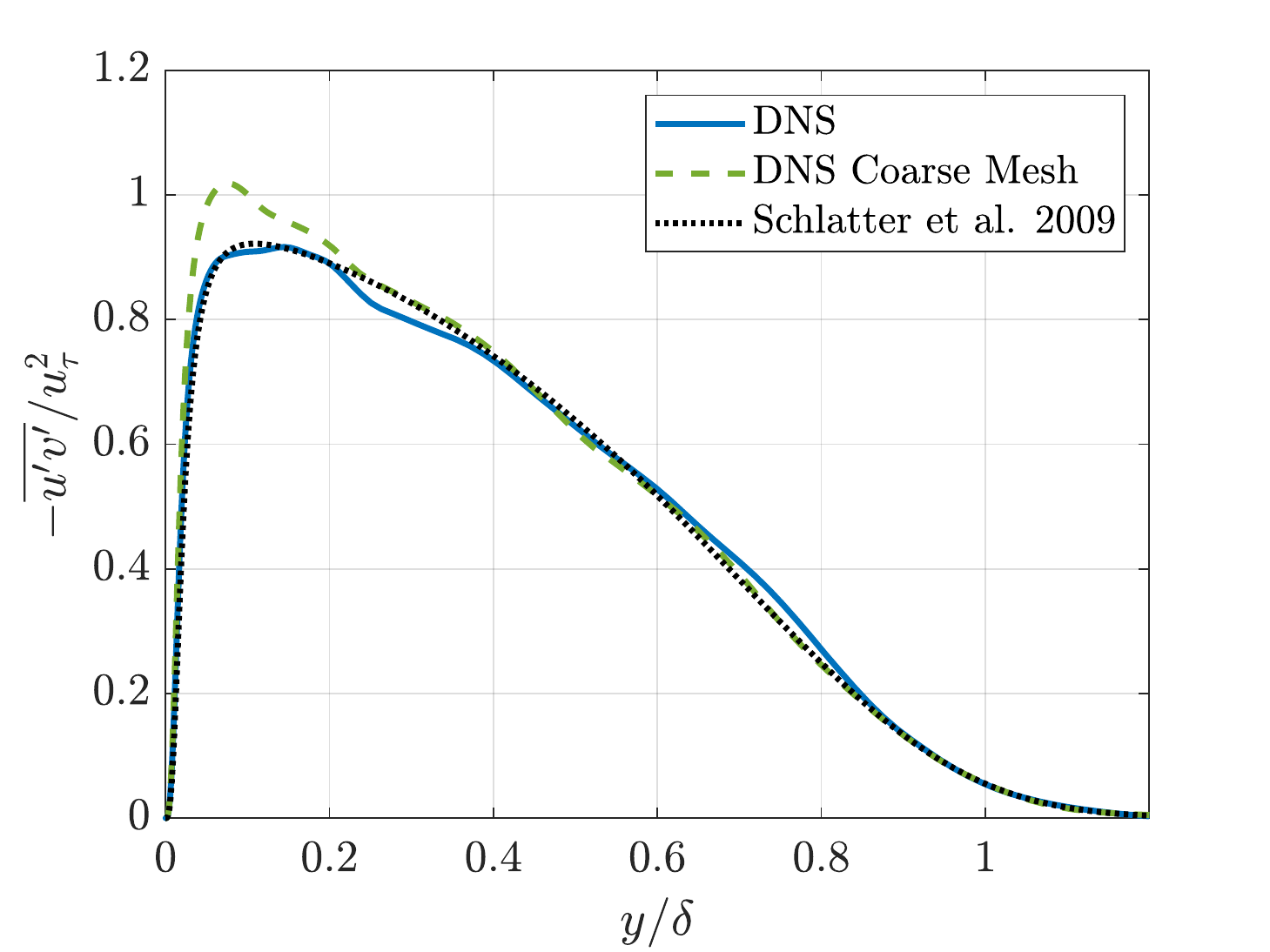}
          \includegraphics[width = 0.33\textwidth]{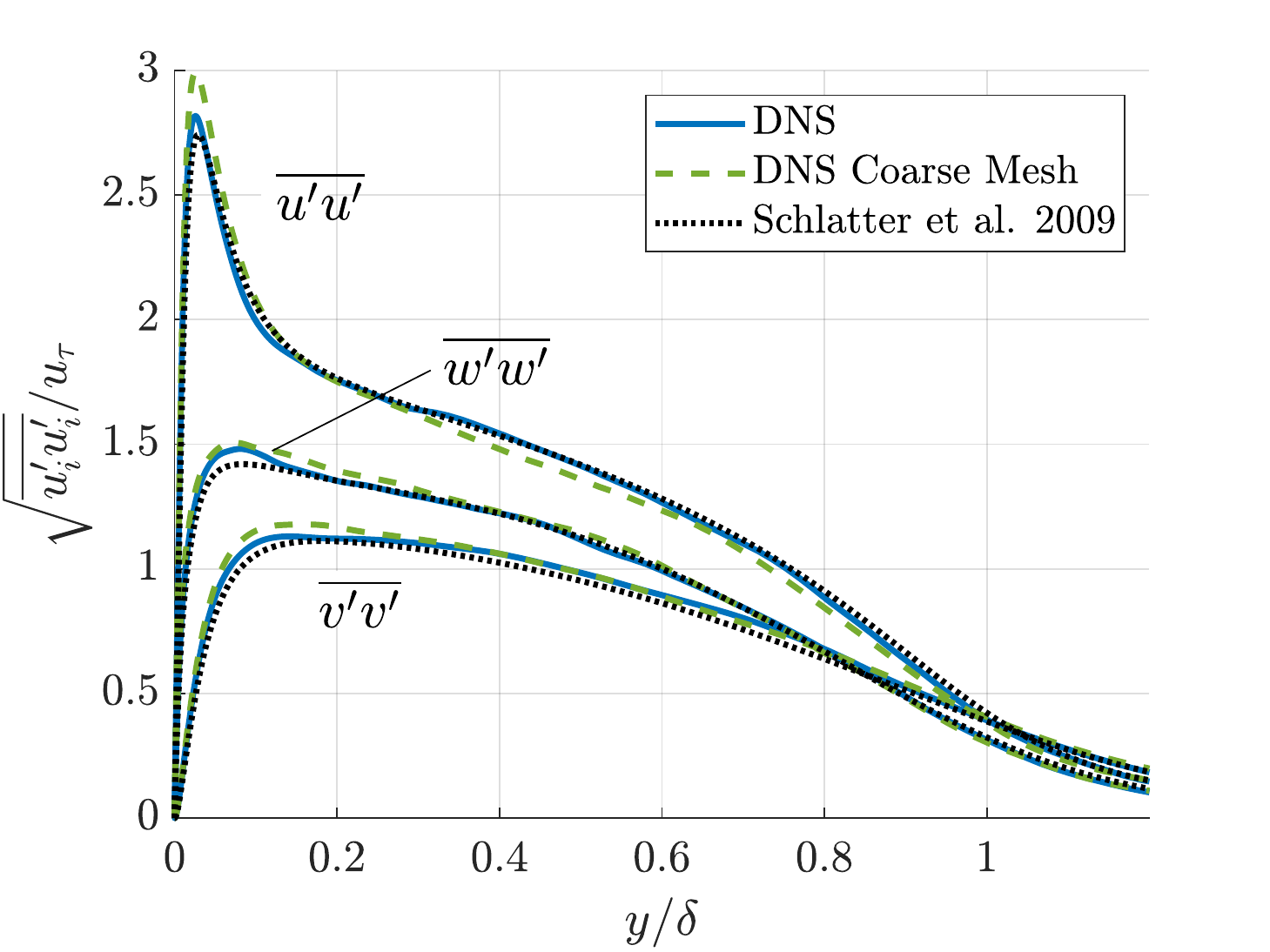}
         \caption{\(Re_\theta = \num{1410}\)}
         \label{fig:GridComp_Re1410}
     \end{subfigure}
        \caption{Profiles of streamwise velocity, Reynolds shear stress, and Reynolds normal stresses at (a) \(Re_\theta=\num{1100}\) and (b) \(Re_\theta=\num{1410}\).}
        \label{fig:GridComp_profiles}
\end{figure}

Profiles of the baseline and coarse mesh DNS are shown in \cref{fig:GridComp_profiles}. 
They are taken at locations corresponding an \(Re_\theta\) of \num{1100} and \num{1410}, and are compared to results from \cite{Jimenez_ZPGDNS} and \cite{Schlatter2009} at those respective locations.
Note that these profile locations correspond to distances of \(7\delta_\tin\) and \(22 \delta_\tin\) downstream of the inlet for \(Re_\theta=\num{1100}\) and \num{1410} respectively. 
Thus, the plots presented in \cref{fig:GridComp_profiles} evaluate the effectiveness of the method to quickly produce realistic turbulence.  

The velocity profiles of both grid resolutions agree very well with the logarithmic law and with the other DNS at both profile locations. 
For the baseline grid, the shear and normal Reynolds stress profiles at \(Re_\theta=\num{1100}\) are not in as good of an agreement with the DNS results of \cite{Jimenez_ZPGDNS} as velocity was.
The normal Reynolds stresses do however show realistic anisotopy, which is spite of the isotropic stresses prescribed at the STG inlet, while the shear Reynolds stress shows the same general behavior as \cite{Jimenez_ZPGDNS}. 
This indicates the formation of correct near-wall turbulence.
At the \(Re_\theta=\num{1410}\) profile location, the Reynolds stresses for the baseline grid agree quite well against the DNS results of \cite{Schlatter2009}, indicating a complete development into ``natural'' turbulence.
Thus, at the earlier profile, the development of the synthetic turbulence is complete enough for accurate first moment statistics (velocity and \(C_f\)), but it not quite enough to match second moment statistics with the other DNS results.

The coarse grid's Reynolds stresses prediction tracks close to the baseline grid results for the outer half of the boundary layer.
Closer to the wall, however, it generally predicts higher stresses.
This trend is present at both profile locations, indicating that while the coarse grid gives first moment statistic predictions that are comparable with the baseline grid, the same cannot be said for its prediction of second moment statistics.


\subsection{Changes to Numerical Dissipation}\label{sec:ResDisComp}
\begin{figure}
     \centering
     \begin{subfigure}[b]{\textwidth}
         \centering
          \includegraphics[width = 0.33\textwidth]{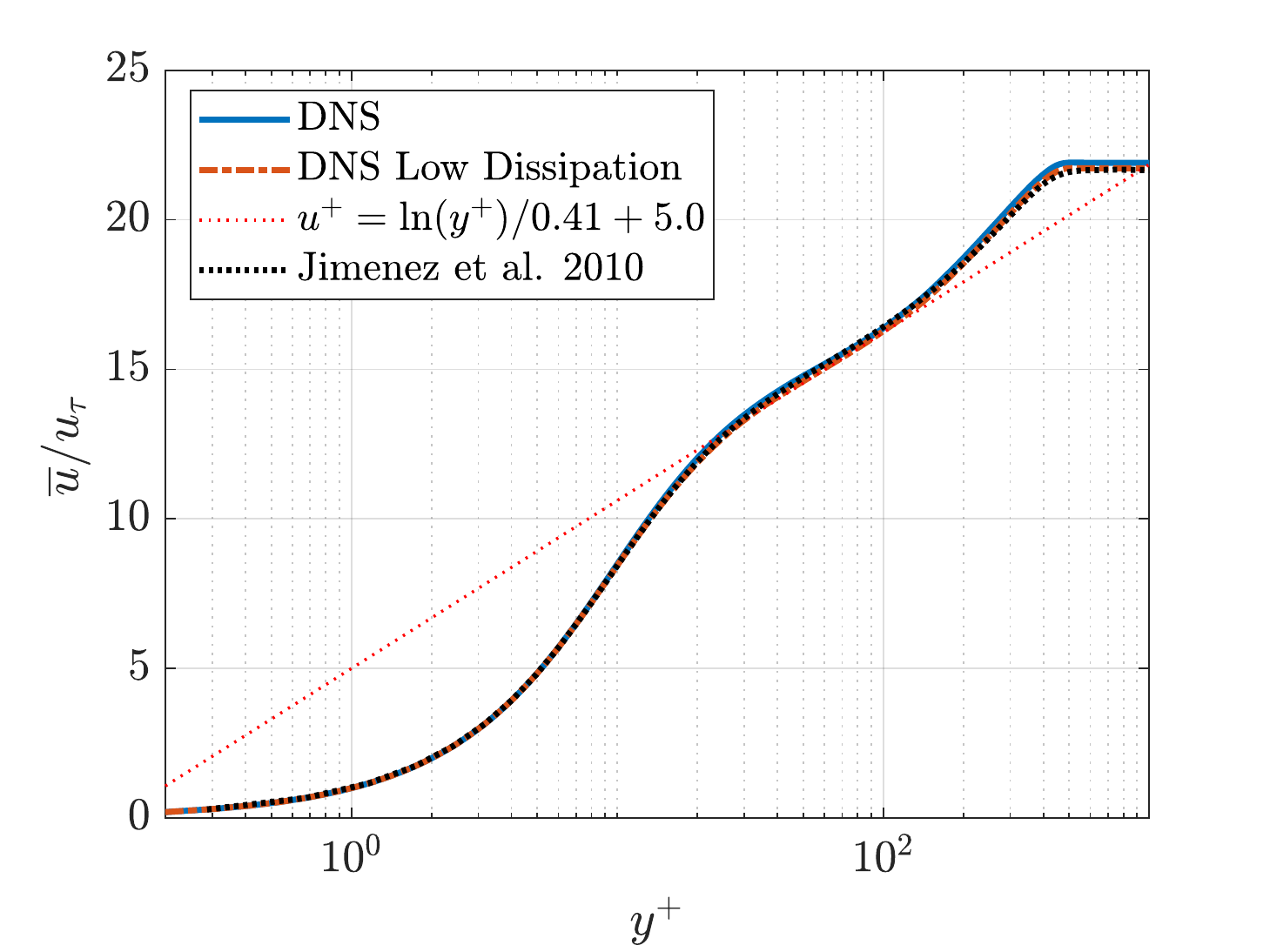}
          \includegraphics[width = 0.33\textwidth]{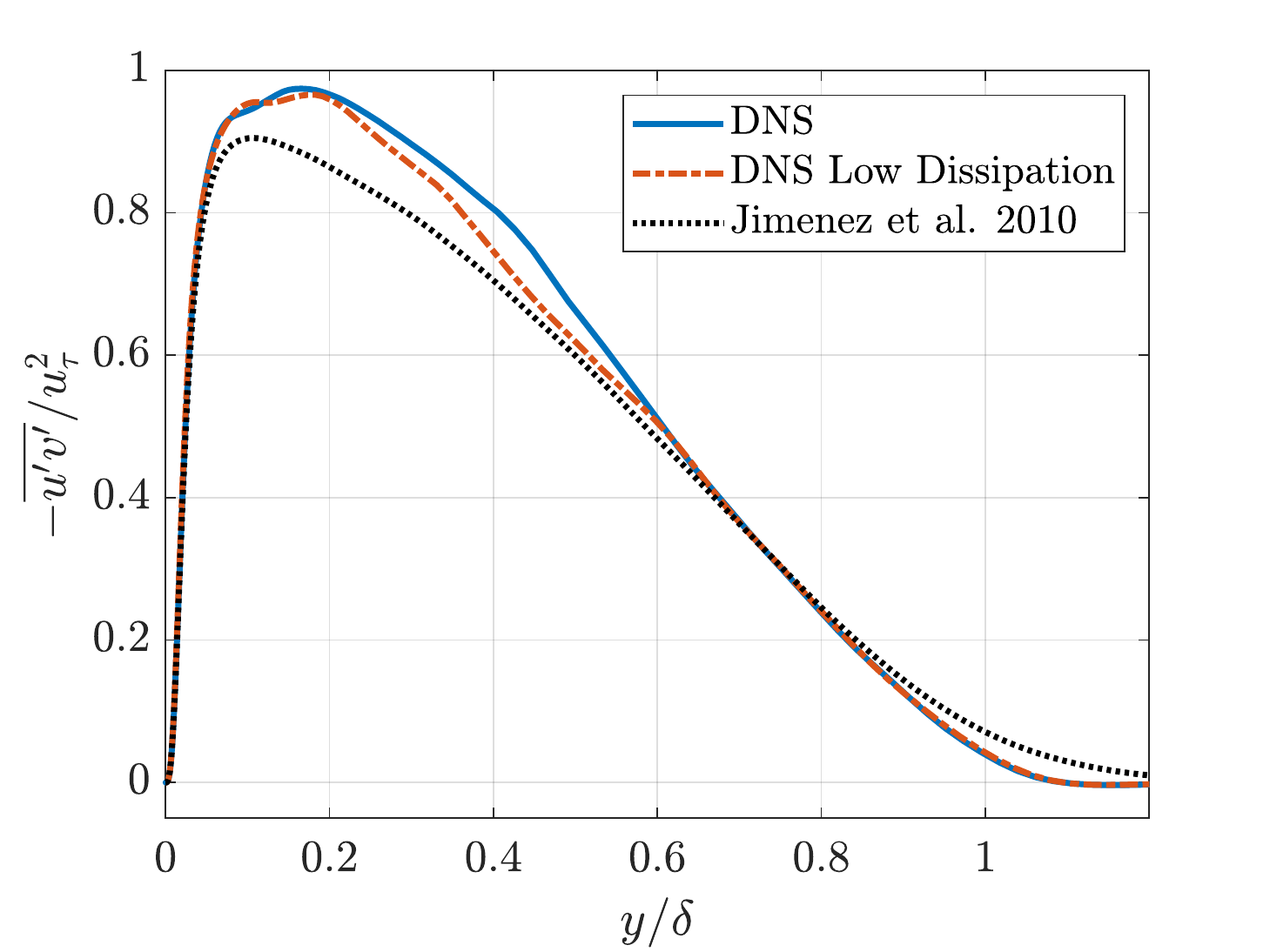}
          \includegraphics[width = 0.33\textwidth]{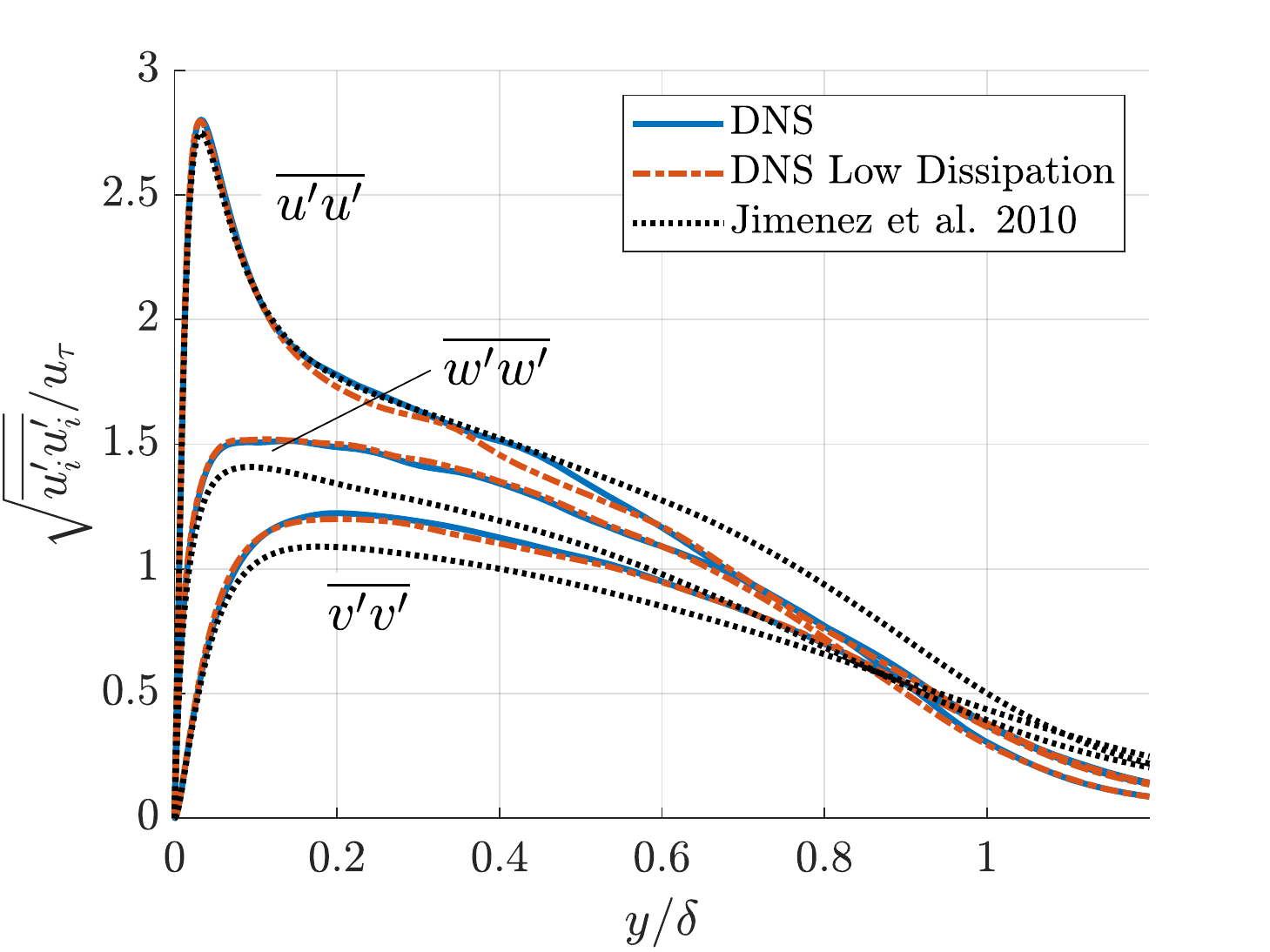}
         \caption{\(Re_\theta = \num{1100}\)}
         \label{fig:DisComp_Re1100}
     \end{subfigure}
     \begin{subfigure}[b]{\textwidth}
         \centering
          \includegraphics[width = 0.33\textwidth]{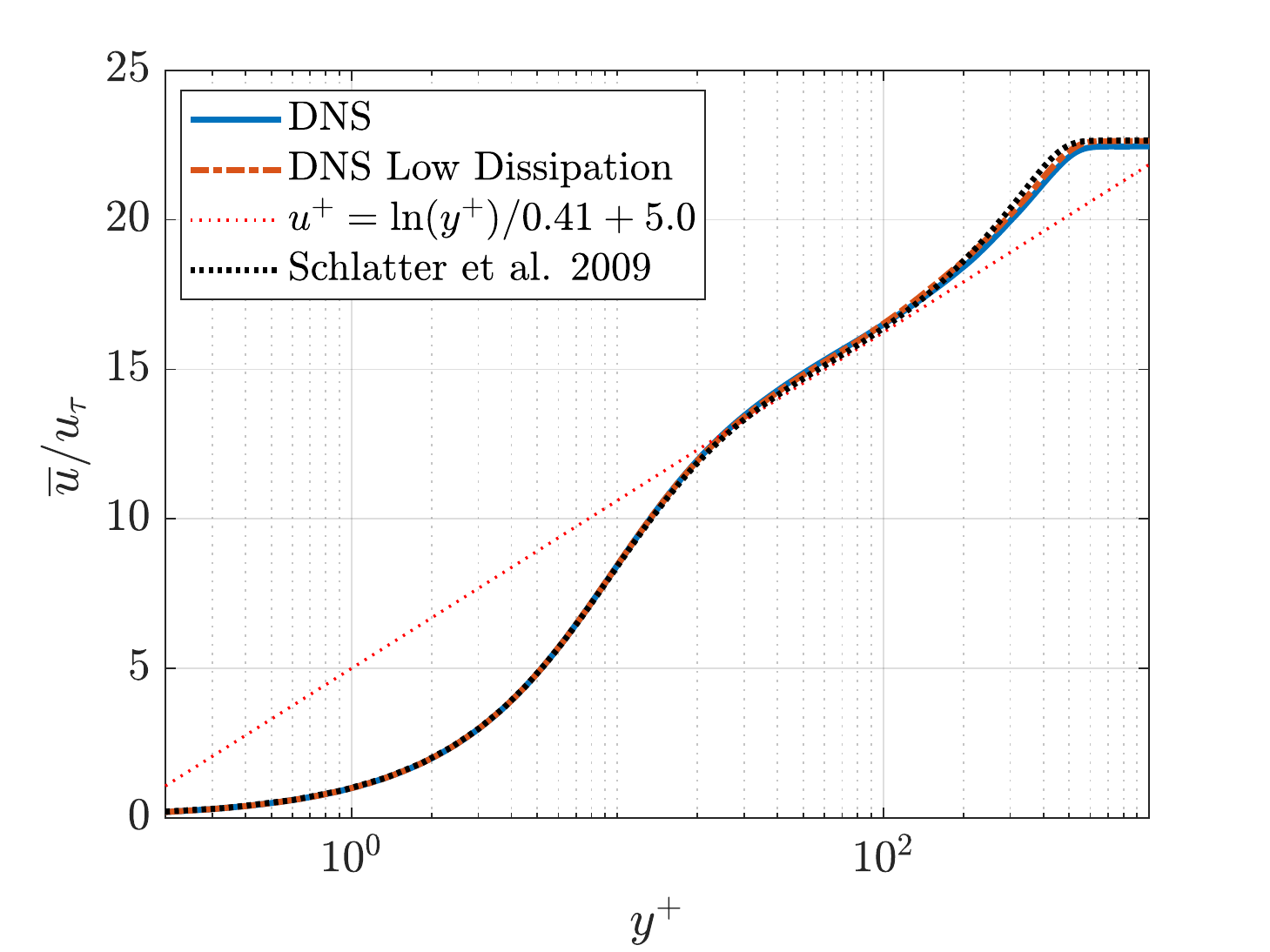}
          \includegraphics[width = 0.33\textwidth]{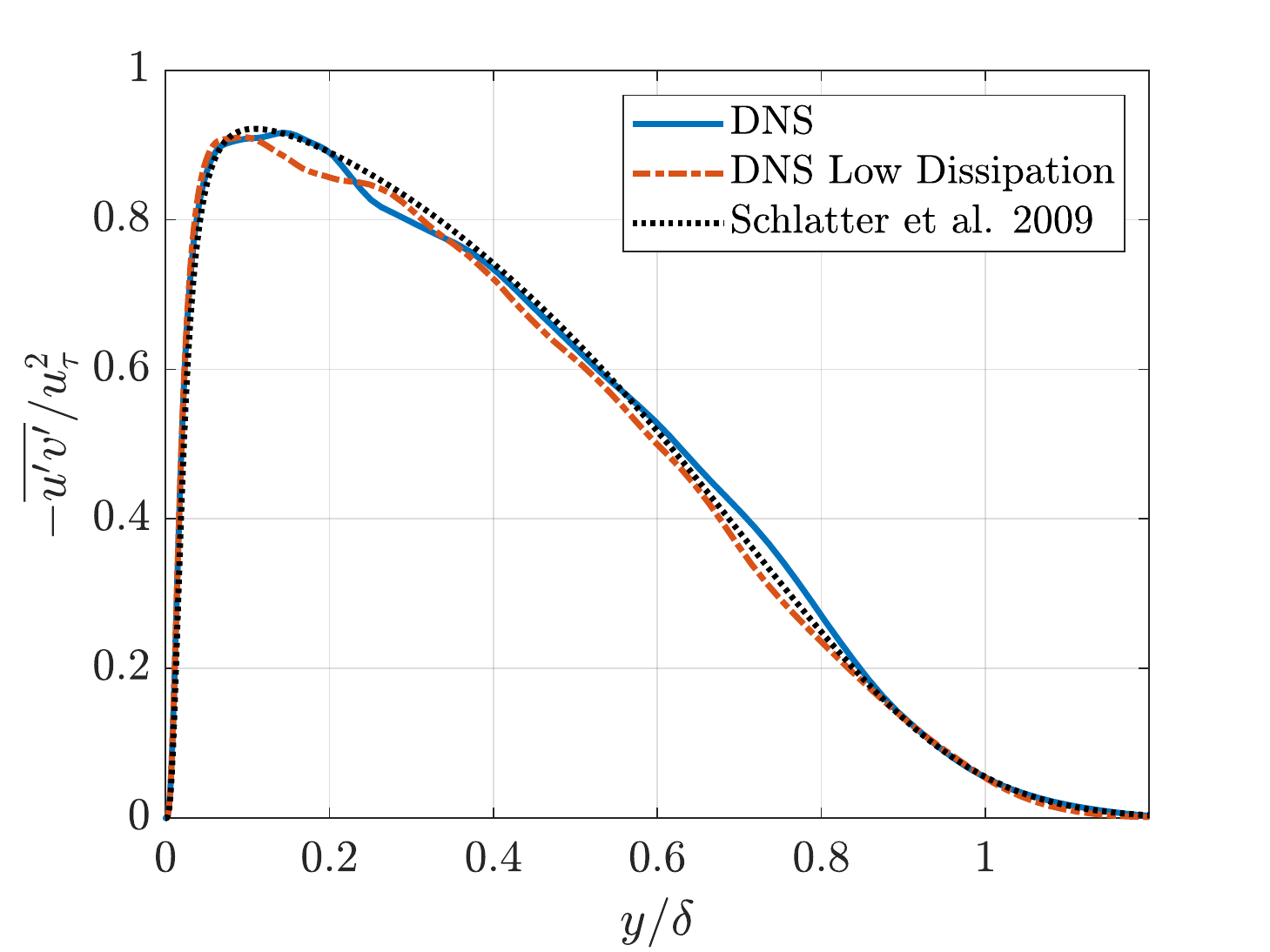}
          \includegraphics[width = 0.33\textwidth]{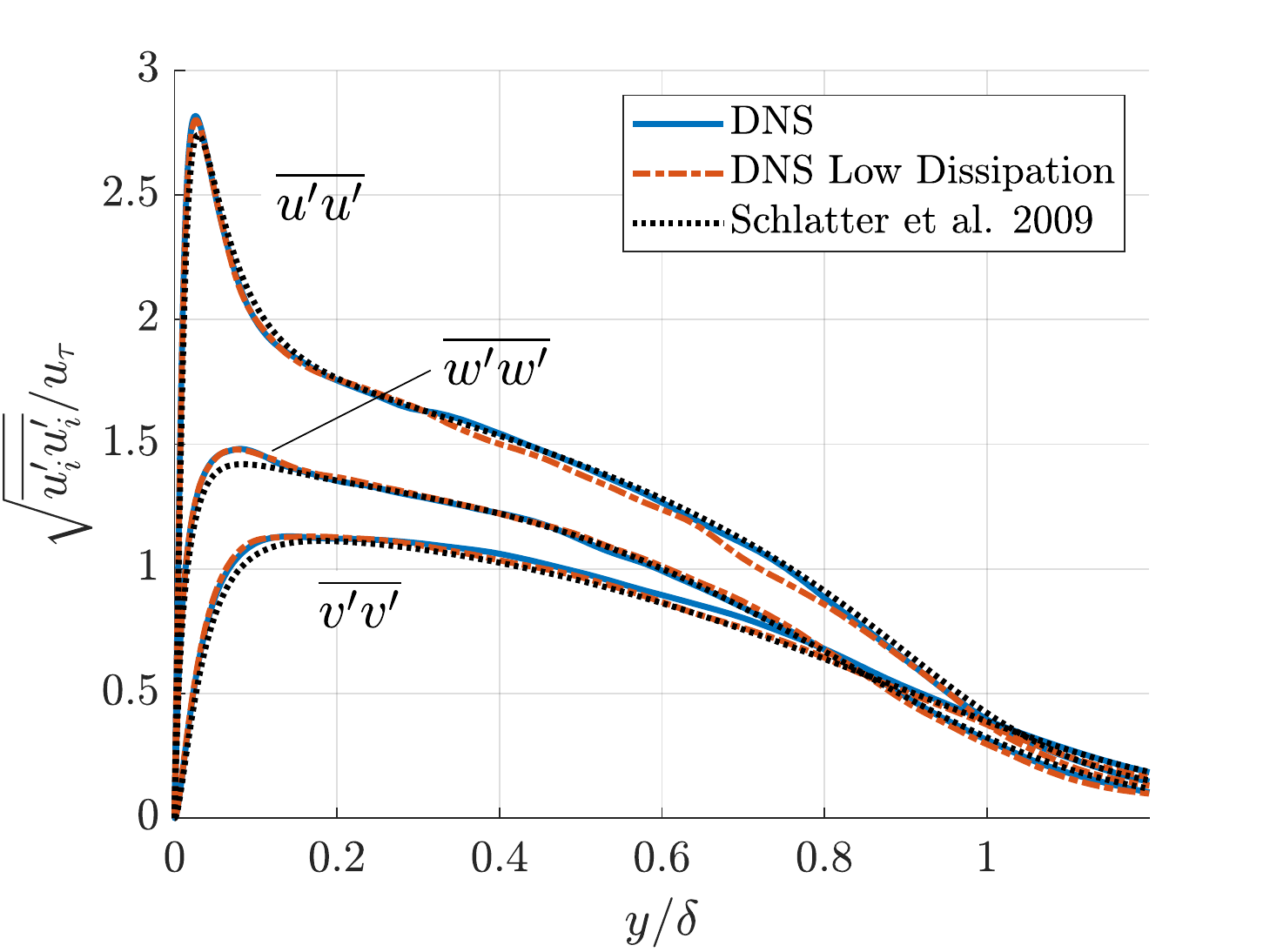}
         \caption{\(Re_\theta = \num{1410}\)}
         \label{fig:DisComp_Re1410}
     \end{subfigure}
        \caption{Profiles of streamwise velocity, Reynolds shear stress, and Reynolds normal stresses at (a) \(Re_\theta=\num{1100}\) and (b) \(Re_\theta=\num{1410}\).}
        \label{fig:DisComp_profiles}
\end{figure}

In addition to mesh resolution, the effect of numerical dissipation on the STG development of this flat plate was also investigated.
To do this, the numerical dissipation parameters in PHASTA were adjusted for the baseline mesh.
Temporal numerical dissipation in PHASTA can be controlled by changing the $\rho_\infty$ parameter of the generalized-$\alpha$ time integrator~\cite{Jansen_GenAlpha_2000}, while 
spatial numerical dissipation can be controlled by adjusting the \(c_1\) parameter of the SUPG stabilization, described in~\cref{sec:AppSUPG}.
The baseline simulations were run with \(\rho_\infty = 0.5\) and \(c_1=8\).
The dashed curves in~\cref{fig:DisComp_Cf,fig:DisComp_profiles} show the results obtained after changing the dissipation parameters to  \(\rho_\infty = 0.75\) and \(c_1=16\).
Differences between the skin friction predictions are 2\% or less and within the scatter due to lack of full convergence of the statistics. 
Differences in the velocity and Reynolds stress profiles are also negligible. 
Consequently, the effects of numerical dissipation at this level of resolution were considered to have a minimal impact on the development of the STG inflow condition into ``natural'' turbulence.

\section{Conclusions}
\label{sec:Conc}

Direct numerical simulation was performed of the turbulent boundary layer over a short flat plate utilizing an STG inflow boundary condition. 
The domain was chosen to be as short as feasible to test the development length of this boundary condition -- how many boundary layer thicknesses are required to reproduce an equilibrium turbulent boundary layer -- as measured by agreement in skin friction coefficient, streamwise velocity, and Reynolds stress profiles with previous numerical and experimental studies.  
A development length of 5-7 boundary layer thicknesses was adequate to match validation data to within 2\% for first order moments and 5--10\% for second order moments.  
Furthermore, this result was shown to be insensitive to stabilization parameters in the stabilized finite element method used.
This, along with comparisons to results from a coarsened mesh, indicate that a stream-wise spacing of 15 plus units, span-wise spacing of 6 plus units, and a maximum wall-normal spacing of 10 plus units is adequate for this flow when discretized with structured tri-linear hexahedra.
\appendix
\section{SUPG Stabilization Formulation} \label{sec:AppSUPG}

Before discussing the formulation of streamline-upwind Petrov-Galerkin (SUPG) in PHASTA, the formulation of the incompressible Navier-Stokes equations used in PHASTA need be defined. 
We define a domain \(\Omega \subset \mathbb{R}^3\) with boundary \(\Gamma\) discretized into elements \(\Omega_e\), where 
\begin{equation}
    \Omega = \bigcup_{e=1}^{n_\text{el}} \Omega_e
\end{equation}

Also, define \(\Gamma_g \subset \Gamma\).
Next, we specify the solution and weight spaces on \(\Omega\).

\begin{equation}
    \mathcal{S}_h^k = \{\vec{v}|\vec{v}(\cdot,t)\in H^1(\Omega) \forall t\in[0,T],\ 
    \vec{v}\vert_{x\in\Omega_e} \in P_k(\Omega_e), \ 
    \vec{v}(\cdot,t) = g \text{ on } \Gamma_g\}
\end{equation}
\begin{equation}
    \mathcal{W}_h^k = \{\vec{w}|\vec{w}(\cdot,t)\in H^1(\Omega) \forall t\in[0,T],\  
    \vec{w}\vert_{x\in\Omega_e} \in P_k(\Omega_e), \ 
    \vec{w}(\cdot,t) = 0 \text{ on } \Gamma_g\}
\end{equation}
\begin{equation}
    \mathcal{P}_h^k = \{p|p(\cdot,t)\in H^1(\Omega) \forall t\in[0,T],\ 
    p\vert_{x\in\Omega_e} \in P_k(\Omega_e) \}
\end{equation}

We seek the solution \(\vec{u} \in \mathcal{S}_h^k\) and \(p \in \mathcal{P}_h^k\) such that a residual \(\mathcal{G}=0\), where
\begin{multline} \label{eq:GNS}
    \mathcal{G}(w_i, q; u_i, p) = 
    \int_\Omega \left \{-q_{,i} u_i 
    + w_i \left[\rho u_{i,t} + \rho u_j u_{i,j} - f_i\right ]
    + w_{i,j} \left[\tau_{ij} - p\delta_{ij}\right ] \right \} \dif \Omega \\
    + \int_\Gamma \left \{ q u_i \hat{n}_i
    + w_i \left[\tau_{ij} - p\delta_{ij}\right ] \hat{n}_j
    \right \} \dif \Gamma
\end{multline}

\noindent
, \(\vec{w}_i \in \mathcal{W}_h^k\) and \(q \in \mathcal{P}_h^k\) are the weight functions for velocity and pressure respectively, \(\tau_{ij}\) is the viscous stress, \(f_i\) are source terms, and \(\hat{n}_i\) is the outward facing normal of \(\Gamma\). 
Note that we have switched to index notation, summation on repeated indices is implied (\(\tau_{ii} = \sum_{i=1}^3 \tau_{ii}\)), and comma notation denotes partial derivatives (\(u_{j,i} = \partial u_j / \partial x_i\)). 
Equation \ref{eq:GNS} is the Galerkin form of the incompressible Navier-Stokes equations, however it is known to be unstable for convection dominated flows (of which our zero-pressure gradient boundary-layer flow is a member of). 
Thus, stabilization is required to achieve solution accuracy.
PHASTA uses SUPG for stabilization, which can be formulated as an addition to the weighted-residual function \(\mathcal{G}\):

\begin{equation} \label{eq:stabGNS}
    \mathcal{G^*}(w_i, q; u_i, p) = \mathcal{G}(w_i, q; u_i, p) 
    + \sum_{e=1}^{n_\text{el}} \int_{\Omega_e} \left \{ 
    \tau_m \left [ u_j w_{i,j} + \frac{q_{,i}}{\rho}\right] \mathcal{L}_i
    + \tau_c w_{j,j} u_{i,i}
    \right \} \dif \Omega_e
\end{equation}

\noindent
where \(\mathcal{L}_i\) is the residual-strong-form of the momentum equations:

\begin{equation}
    \mathcal{L}_i = \rho u_{i,t} + \rho u_j u_{i,j} + p_{,i} - \tau_{ij,j} - f_i
\end{equation}

The terms in equation \ref{eq:stabGNS} multiplying \(\tau_m\) and \(\tau_c\) stabilize continuity and momentum respectively.

\begin{equation}
    \tau_m = \left [
    \left(\frac{2 c_1}{\Delta t} \right)^2 + u_i u_j g_{ij} + 
    c_2 f \nu^2 g_{ij} g_{ij} \right ] ^{-1/2}
\end{equation}

\begin{equation}
    \tau_c = \frac{ c_3 \rho}
    {8 c_1 g_{ii} \tau_m}
\end{equation}

\noindent
where \(c_1\), \(c_2\), and \(c_3\) are adjustable coefficients, \(g_{ij}\) is the metric tensor for each element (\(\xi_{k,i}\xi_{k,j}\)), and \(f\) is a correction factor based on basis function order. 
For first-order basis functions, \(f=36\). For the simulations run in this paper, \(c_2\) and \(c_3\) were set to 1, while \(c_1\) was adjusted as described in Section \ref{sec:Results}.

\section*{Acknowledgements}
This work was supported by the National Science Foundation, Chemical, Bioengineering, Environmental and Transport Systems grant CBET-1710670 and by the National Aeronautics and Space Administration, Transformational Tools and Technologies grant 80NSSC18M0147, both to the University of Colorado Boulder. 
Computational resources were utilized at the NASA High-End Computing (HEC) Program through the NASA Advanced Supercomputing (NAS) Division at Ames Research Center and at the Argonne Leadership Computing Facility (ALCF), which is a DOE Office of Science User Facility supported under Contract DE-AC02-06CH11357. 
Finally, the authors thank Drs. P.R. Spalart, and M.K. Strelets for the helpful insight and communications regarding the problem setup, the analysis of the flow, and the synthetic turbulence generation method.

\bibliographystyle{unsrt}  
\bibliography{biblio}

\end{document}